\documentclass[aps,pra,twocolumn,superscriptaddress]{revtex4-1}

\usepackage{amsmath}
\usepackage{amssymb}
\usepackage{epsfig}
\usepackage{color}
\usepackage{soul}
\usepackage{epstopdf}
\usepackage{graphicx}
\usepackage[usenames,dvipsnames]{xcolor}

\graphicspath{{.}{./figs/}}

\newcommand{\rr}{{\boldsymbol r}}
\newcommand{\sub}[1]{{\textrm{#1}}}

\providecommand{\ket}[1]{\ensuremath{\left\vert #1\right\rangle}}
\providecommand{\bra}[1]{\ensuremath{\left\langle #1\right\vert}}

\begin{document}

\title{Sensing microwave photons with a Bose-Einstein condensate}

\author{Orsolya K\'{a}lm\'{a}n}
\author{Peter Domokos}
\affiliation{
Institute for Solid State Physics and Optics, Wigner Research Centre for Physics, Hungarian Academy of Sciences, 
P.O. Box 49, H-1525
Budapest, Hungary}

%\date{\today}
\begin{abstract}
We consider the interaction of a magnetically trapped Bose-Einstein condensate of Rubidium atoms with the stationary microwave radiation field sustained by a coplanar waveguide resonator. This coupling allows for the measurement of the magnetic field of the resonator by means of counting the atoms that fall out of the condensate due to hyperfine transitions to non-trapped states. We determine the quantum efficiency of this detection scheme and show that weak  microwave fields at the single-photon level can be sensed with an integration time on the order of a second.    
\end{abstract}
\maketitle

\section{Introduction}
\label{sec:intro}

Sensing extreme low-intensity radiation fields in the quantum regime requires high-efficiency, low-noise detectors. In the case of the optical wavelength range, single-photon detectors are widely available. The thermal noise at optical frequencies is naturally suppressed at room temperature. Moreover, the electric dipole transitions in atoms or in semiconductors give rise to large coupling to the electromagnetic field and, ultimately, to high quantum efficiency. By contrast to the optical case, photon counting of microwave radiation is still a formidable task.  It is possible to count microwave photons of a resonator, for example,  by means of Ramsey interferometry with highly-excited circular Rydberg atoms \cite{Gleyzes2007}. Analogous schemes have been demonstrated for  microwave waveguide resonators integrated in a circuit with strongly coupled superconducting non-linear elements, such as e.g. Josephson junctions \cite{JBAreview2009}.  In these chip-based circuits the quantization of flux and charge leads to quantized electric and magnetic fields at microwave frequencies, in accordance with the longitudinal dimension of the device \cite{Girvin_LesHouches}. However, the resolved detection of photons based on the same principle as that of commonly used optical detectors, i.e., the absorption by  a ground-state material probe, is missing in the microwave regime.

In the microwave frequency range,  the naturally occurring transitions in  material quantum systems are magnetic ones, which are weaker than the electric dipole transitions typically by a factor of the fine structure constant $\alpha=1/137$. One promising candidate is the spin degree of freedom of point defects in nanocrystals, which provides for a portable, well localised probe of the field \cite{Taylor2008,Li2015}. Alternatively, hyperfine transitions of atoms in the electronic ground state can be considered \cite{Wildermuth2005}. In this approach, the quantum efficiency of the detection can benefit from the collective enhancement gained by using a degenerate atomic cloud, i.e., a Bose-Einstein condensate (BEC) of trapped atoms. The price to pay is that the condensate is not so well localised as a single spin or atom, nevertheless, it is still orders of magnitude smaller than the corresponding wavelength of the radiation. The trapped BEC is an ideal probe of weak external fields due to the fact that all their relevant degrees of freedom can be controlled with unprecedented precision \cite{fortagh07,Folman_review}. That is, it approaches closely the ultimate quantum-noise limit requested from a detector.

In this paper we evaluate the detection capabilities of a BEC in the microwave regime.  In particular, we investigate whether a microwave field at the single-photon level can be sensed and translated to a detectable signal by means of the atomlaser scheme \cite{Ottl_05,Kohl2007} for magnetic noise measurement \cite{Kalman_12,Gunther_15,Kalman2016}.  It has been first proposed in Ref. \cite{Verdu2009} that strong-coupling between ultracold  atoms and the magnetic field of a waveguide resonator can be achieved. Recently, the near-field microwave radiation of a coplanar waveguide resonator (CPW) has been successfully coupled to ultracold atoms  \cite{Fortagh2017}. In this experiment the light-shift of a hyperfine transition induced by the  strongly driven resonator mode in the large photon number regime was detected by Ramsey interferometry and the possibility of coherent control of the hyperfine states was demonstrated by directly observing resonant Rabi oscillations. Here we consider the opposite, weak field limit and restrict our aim at the detection of a feeble radiation field. To this end, we resort to the atomlaser detection method which relies on counting single atoms outcoupled  by the measured field and falling out of the trap due to gravity \cite{Ottl_05,Kalman2016}.

The paper is organized as follows. In Sec.~\ref{sec:sys} we consider a feasible setup where the condensate is situated at the antinode of the magnetic field under the microwave resonator, so that the outcoupled atoms can fall out of the trap due to gravity. Sec.~\ref{sec:BEC} is devoted to the description of the atomlaser scheme, where an appropriate hyperfine transition is driven by a microwave-frequency magnetic field, and the density of outcoupled atoms is determined using the results of \cite{Kalman2016}. Sec.~\ref{sec:CPW} overviews the geometry and the description of the coplanar waveguide resonator.  In Sec.~\ref{sec:quant_eff} we consider the process of sensing the magnetic field of a single photon in the resonator. First, in Sec.~\ref{subsec:num_outcoupl} we give an approximation for the magnetic field when there is, on average, a single photon in the resonator, and present an analytical formula for the number of outcoupled atoms. Then, in Sec.~\ref{subsec:num_at} we consider typical parameters for the BEC and the CPW for our setup and estimate how many outcoupled atoms would be found per a single microwave photon in a unit of time assuming perfect ionization detection of the atoms. We conclude in Sec.~\ref{sec:concl}.

\section{The setup}
\label{sec:sys}

\begin{figure}[thb]
\includegraphics[width=0.99\columnwidth]{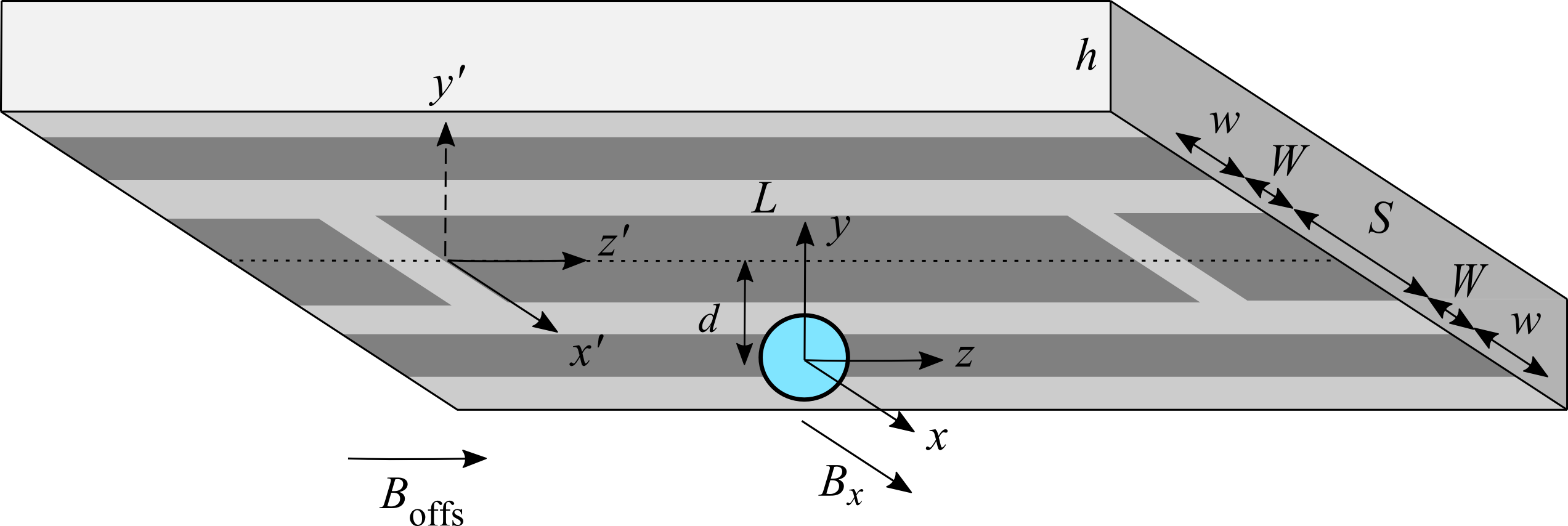}
\caption{(color online) Sketch of the the setup to sense the magnetic field of a CPW with a BEC.}
\label{Fig1}
\end{figure}

In order to achieve maximal coupling between the BEC and the CPW, the ultracold rubidium atom cloud is situated at the center of central conductor of the waveguide resonator, as shown in Fig.~\ref{Fig1}, with the CPW above the condensate so that the outcoupled atoms can fall out and once spatially separated from the trap, be detected and counted with single-atom precision \cite{Stibor2010}.  

%%%%%%%%%%%%%%%%%%%%%%%%%%%%%%%%%%%%%%%%%%%%%%%%%%%%%%%%%%%%%%%%%%%%%%%%%%%%%%%% 
%%%%%%%%%%%%%%%%%%%%%%%%%%%%%%%%%%%%%%%%%%%%%%%%%%%%%%%%%%%%%%%%%%%%%%%%%%%%%%%%

\subsection{The Bose-Einstein condensate}
\label{sec:BEC}

We consider ultracold ${}^{87}$Rb atoms prepared in the hyperfine manifold $F=2$ of the ground state $5\,^{2}\mathrm{S}_{1/2}$, placed in an external magnetic field $\mathbf{B}(\rr)$ in the presence of gravity, with gravitational acceleration $g_{\mathrm{gr}}$. The total atomic angular momentum $\hat{\mathbf F}$ interacts with the magnetic field according to the Zeeman term $H_{Z} = g_{F} \, \mu_{\mathrm B} \hat{\mathbf F}{\mathbf B}(\rr)$, where $\mu_{\mathrm B} = e\hbar/2 m_{e}$ is the Bohr magneton, and $g_{F}$ is the Land\'{e} factor.  The dominant component of the magnetic field $\mathbf{B(\rr)}$ is a homogeneous offset field $B_{\rm offs}$ pointing along the $z$ direction. The eigenstates of the spin component $\hat F_{z}$, labelled by $m_F=-2,-1,0,1,2$, are well separated by the Zeeman shift $\hbar\omega_{0}=\mu_{\mathrm{B}}B_{\mathrm{offs}}/2$, the $m_{F}=2$ level being the highest in energy (due to the fact that $g_{F=2}=1/2$). The inhomogeneous component of the magnetic field $\mathbf{B}(\rr)$ creates a harmonic trapping potential around the minimum of the total magnetic field in which we assume atoms to be confined in the low-field seeking state $\ket{2,1}$. To a good approximation, $\mathbf{B}(\rr)$ gives rise to the static potential $V_{\ket{2,1}}(\rr)=\hbar\omega_{\mathrm{L}}+V_{\mathrm{T}}(\rr)$, where $\omega_{\mathrm{L}}=\frac{1}{\hbar}\left[\hbar\Omega+\hbar\omega_{0}+\frac{Mg_{\mathrm{gr}}^{2}}{2\omega_{y}^{2}}\right]$ is the  Larmor frequency in the presence of gravity, $\hbar\Omega=\hbar^{2}A_{\mathrm{hfs}}\approx6.8347$ being the hyperfine splitting of the $5\,^{2}\mathrm{S}_{1/2}$ state, and  $V_{\mathrm T}(\rr)=\frac{M}{2}\left[\omega_{x}^{2}x^{2}+ \omega_{y}^{2}y^{2}+\omega_{z}^{2}z^{2}\right]$ is the harmonic trapping  potential with $\omega_{x}$, $\omega_{y}$ and $\omega_{z}$ being the trap frequencies in the $x$, $y$, and $z$ directions, respectively, and $M$ being the atomic mass. Note that the origin of the coordinate system does not coincide with the minimum of the magnetic field but is displaced by a significant gravitational sag $y_{0}=-g_{\mathrm{gr}}/\omega_{y}^{2}$. 

In the magnetically trapped $\ket{2,1}$ state,  we assume a pure Bose-Einstein condensate (BEC) described by the second quantized field operator $\hat{\Psi}_{\ket{2,1}}(\rr,t)= \sqrt{N_{0}} \Phi_{\mathrm{BEC}}(\rr)e^{-i(\omega_{\mathrm{L}}+\mu/\hbar)t}$, where the wavefunction $\Phi_{\mathrm{BEC}}$ is the stationary solution of the Gross-Pitaevskii equation with chemical potential $\mu$ and atom number $N_{0}$. 

We have previously considered the outcoupling of BEC atoms by an external, spatially homogeneous, time-varying magnetic field polarized in the $x$ direction, with a monochromatic carrier frequency in the radio-frequency domain, which could quasi-resonantly induce magnetic dipole transitions from the trapped state $\ket{1,-1}$ into the untrapped state $\ket{1,0}$ \cite{Kalman2016}. Here we couple the BEC to a microwave cavity in a similar manner, based on the fact that magnetic dipole transitions among the $F=2$ and $F=1$ manifolds are also possible. The homogeneous offset magnetic field splits the magnetic sublevels of the $F=1$ manifold as well, the splitting between consecutive levels also being $\hbar \omega_{0}$. By decomposing the total magnetic moment of the atom as $\boldsymbol{\mu}=\boldsymbol{\mu}_{L}+\boldsymbol{\mu}_{S}+\boldsymbol{\mu}_{I}$, it can be shown that in the $5\,^{2}\mathrm{S}_{1/2}$ ground state only the $\hat{S}_{+}$ and $\hat{S}_{-}$ operators have nonvanishing matrix elements between the magnetic sublevels of the different $F$ manifolds. (This can be seen by expressing the basis states $\left\lbrace \ket{F,m_{F}},\, F=1,2,\, m_{F}=-F,...,F \right\rbrace$ common to the operators $\hat{\mathbf{S}}^{2},\hat{\mathbf{I}}^{2},\hat{\mathbf{F}}^{2},\hat{F}_{z}$ with the basis states $\left\lbrace \ket{m_{S},m_{I}},\, m_{S}=-1/2,1/2,\, m_{I}=-3/2,...,3/2 \right\rbrace$ common to the operators $\hat{\mathbf{S}}^{2},\hat{\mathbf{I}}^{2},\hat{S}_{z},\hat{I}_{z}$.) Here we consider transitions between hyperfine states $\ket{2,1}$ and $\ket{1,0}$ and therefore focus on the matrix element $\bra{1,0}\hat{S}_{-}\ket{2,1}=-\sqrt{3}\hbar/2\sqrt{2}$. 

Atoms in the state $\ket{1,0}$, which can be described by the field operator $\hat{\Psi}_{\ket{1,0}}(\rr,t)$ are not affected by the offset field and the inhomogeneous trapping field, they move under the influence of gravity and the mean-field potential $N_{0}g_{s}\Phi_{\mathrm{BEC}}^{2}(\rr)$, with $g_{s}=4\pi\hbar^{2}a_{s}/M$, and scattering length $a_{s}$ ($a_{s}$=5.4 nm for ${}^{87}$Rb). We consider a setup where at the location of the BEC the microwave magnetic field is quasi-homogeneous in the $x$ direction, so that the perturbation caused by the magnetic field can be written as $V_{I}=g_{S}\mu_{\mathrm{B}} B_{x}(t) \hat{S}_{x}/\hbar$, where $\hat{S}_{x}=(\hat{S}_{+}+\hat{S}_{-})/2$, and $B_{x}(t)=B_{x}\cos(\omega_{\mathrm{CPW}}t)$. We assume that initially no atoms populate the $\ket{1,0}$ state. To leading order in  the small quantum field amplitude $\hat{\Psi}_{\ket{1,0}}$, the equation of motion in rotating-wave approximation reads
\begin{multline}
\!\!\!i\hbar\frac{\partial}{\partial t}\hat{\Psi}_{\ket{1,0}}\!=\! 
\left[-\frac{\hbar^{2}\nabla^{2}}{2M}+Mgy+
N_{0}g_{s}\!\Phi_{\mathrm{BEC}}^{2}(\rr)\right]
\hat{\Psi}_{\ket{1,0}} \\ 
-\hbar \, \eta \, \Phi_{\mathrm{BEC}}(\rr) \, e^{i\Delta\cdot t}\,
, \label{time_ev} 
\end{multline}
where $\eta=\sqrt{3}\mu_{\mathrm{B}} B_{x} \sqrt{N_{0}}/4\sqrt{2}\hbar$ and $\Delta=\omega_{\mathrm{CPW}}-\omega_{\sub L}-\mu/\hbar$ is the detuning of the microwave frequency from the transition frequency at the trap center. Here we considered the BEC as an undepleted reservoir, i.e., the quantum fluctuation $\delta\hat{\Psi}_{\ket{2,1}}$ was neglected in comparison with $\Phi_{\mathrm{BEC}}$ as it corresponds to a second-order process. Furthermore, we assumed that the quantum field components in the $\ket{1,m_F\neq0}$, or the $\ket{2,m_F\neq1}$ sublevels are also negligible when the $\ket{2,1}$$\rightarrow$$\ket{1,0}$ transition is on resonance with the microwave frequency as they are detuned by $\Delta\omega\geq\hbar\omega_{0}$. Finally, we can neglect the transitions $\ket{1,0}$$\rightarrow$$\ket{2,1}$ back to the condensate, originating from the matrix element $\bra{2,1}\hat{S}_{+}\ket{1,0}\neq0$, because the atoms fall out of the trap before completing a Rabi cycle. Within these approximations, the dynamics of the outcoupled field $\hat{\Psi}_{\ket{1,0}}(\rr,t)$ is decoupled from the other Zeeman states. 

In Ref.~\onlinecite{Kalman2016} we gave the solution to the partial differential equation that was similar in form to Eq.~(\ref{time_ev}) with the approximation that the the mean-field potential $N_{0}g_{s}\!\Phi^{2}_{\mathrm{BEC}}(\rr)$ is negligible compared to the gravitational potential. We used the Green function of the corresponding free-fall problem to determine the density of outcoupled  atoms per unit time
\begin{equation}
N\!\left(\Delta,\rr\right)= \left|\frac{\hbar\eta}{Mgl_{0}}\right|^{\!2}
\int\limits_{-\infty}^{\infty}\!d\omega\,D(\omega-\Delta,\rr)\,S(\omega)
\label{N_Delta}
\end{equation}
 and the so-called spectral resolution function $D(\omega-\Delta,\rr)$ of the BEC as a measuring device (see App.~\ref{AppC}), where $S(\omega)$ is the power spectrum of the outcoupling magnetic field, and $l_{0}=\left(\hbar^{2}/2M^{2}g\right)^{1/3}$ is the natural length of the Airy function (free fall problem). The integral over $\omega$ represents the fact that due to the presence of gravity, frequency components that are larger (smaller) than $\Delta$ (the detuning of the carrier frequency of the time-varying magnetic field from the transition frequency of the atoms at the BEC center), can also couple out atoms from the cloud since they may be resonant at $y<0$ ($y>0$). $D(\omega-\Delta,\rr)$ has the dimension of 1/volume and is dependent on the geometry of the BEC cloud. For a spherical BEC that is several times larger than $l_{0}$ in every direction, it is symmetric and has a maximum at $\omega-\Delta\approx 0$ \cite{Kalman2016}.

Here we focus on outcoupling by a monochromatic time-varying magnetic field, which corresponds to $S(\omega)\sim\delta(0)$. The density of outcoupled atoms in this case is given by 
\begin{equation}
N(\Delta,\rr)\!=\!\left<\hat{\Psi}_{\ket{1,0}}^{\dagger}\!(\rr)\,\hat{\Psi}_{\ket{1,0}}\!(\rr)\!\right>\!=\!
\left|\frac{\hbar\eta}{Mgl_{0}}\right|^{2}\!\!D(0-\Delta,\rr).
\label{N_D}
\end{equation} 
$N(\Delta,\rr)$ is maximum for $\Delta\approx 0$, i.e., when the detuning of the frequency of the outcoupling magnetic field from the transition frequency at the center of the BEC is zero.

\subsection{Coplanar waveguide resonator}
\label{sec:CPW}

We consider a (superconducting) coplanar waveguide resonator (CPW) of length $L$ with a central conductor strip of width $S$ and two ground electrodes of width $w$ at a distance $W$ from the central conductor situated on a substrate of thickness $h$ and relative permittivity $\varepsilon_{\mathrm{r}}$ (see Fig.~\ref{Fig1}). In order to effectively couple the CPW to the hyperfine transition $\ket{2,1}\rightarrow\ket{1,0}$ of the BEC atoms, the microwave frequency of the resonator should be resonant with the transition frequency of the atoms at the center of the BEC, i.e., $\omega_{\mathrm{CPW}}\approx \omega_{\mathrm{L}}+\mu/\hbar$. Due to the effect of gravity, the transition frequencies at the top and bottom of the atom cloud are detuned by $-Mga/\hbar$ and $Mga/\hbar$ from $\omega_{\mathrm{CPW}}$, respectively, with $a$ being the semi-axis of the condensate in the direction of gravity. As long as the corresponding bandwidth is  larger than the linewidth of the CPW resonator mode, the full power spectrum of the field mode contributes to the outcoupling of the atoms. In the case we will consider, there is an order of magnitude difference, which justifies the monochromatic approximation leading form  Eq.~(\ref{N_Delta}) to Eq.~(\ref{N_D}).

The CPW's mode wavelength corresponding to $\omega_{\mathrm{CPW}}$ is $\lambda_{g}=\lambda/\sqrt{\varepsilon_{\text{eff}}}=2\pi c/(\omega_{\mathrm{CPW}}\sqrt{\varepsilon_{\mathrm{eff}}})$ where $c$ is the speed of light in air, and $\varepsilon_{\text{eff}}$ is the effective dielectric constant, which can be aproximated by analytical methods using the conformal mapping technique \cite{Simons_CPWbook} c.f. App.~\ref{AppA}. Here we consider a CPW of length $L=\lambda_{g}/2$, and assume that only the fundamental longitudinal mode is excited, for which the components of the magnetic field can be analytically determined using the quasi-static approximation \cite{Simons_CPWbook}, c.f. App.~\ref{AppB}. Using Eqs.~(\ref{Bxyz}) we can estimate the mode volume $V_{c}=\int_{V}d^{3}\rr |\mathbf{B}(\rr\rq)|^{2}/|B_{\mathrm{max}}|^{2}$ of the CPW cavity for the different transverse modes $n$. It can be shown that the mode volume scales inversely with $n$, it is the largest for the lowest transverse mode ($n=1$) for which it can be approximated as $V_{c}\approx Lb^{2}/\pi$. In what follows we will restrict ourselves to the case when only the lowest transverse mode is excited in the CPW.

\section{The quantum efficiency of the sensing process}
\label{sec:quant_eff}

\subsection{The number of outcoupled atoms}
\label{subsec:num_outcoupl}

In order to determine the number of outcoupled atoms per a single microwave cavity photon, we estimate the maximum of the inhomogeneous magnetic field corresponding to one energy quantum $\hbar\omega_{\mathrm{CPW}}$ in the microwave cavity by
\begin{equation}
\left|B_{\mathrm{max}}\right|\!=\!\sqrt{\frac{2\mu_{0}\hbar\omega_{\mathrm{CPW}}}{V_{c}}},
\label{B_est}
\end{equation}
since when the magnetic field is maximal (in time), then all of the photon's energy is carried by the magnetic field, c.f. App.~\ref{AppB}. 

Alternatively, one can estimate $\left|B_{\mathrm{max}}\right|$ based on the circuit quantum electrodynamical formulation \cite{Girvin_LesHouches}, where the amplitude of the voltage operator is given by $V_{0}=\sqrt{\hbar\omega_{\mathrm{CPW}}/C_{\mathrm{CPW}}}$, with $C_{\mathrm{CPW}}$ being the total capacitance of the cavity. Substituting this into Eqs.~(\ref{Bxyz}) and using Eq.~(\ref{c_CPW}) the maximum of the magnetic field can be written as
\begin{equation}
\left|B^{\mathrm{CQED}}_{\mathrm{max}}\right|=\sqrt{\frac{2}{\kappa_{0}}} s_{1}\sqrt{\frac{2\mu_{0}\hbar\omega_{\mathrm{CPW}}}{L b^{2}}}.
\end{equation}
We will show that for the physical parameters to be considered in our calculations, the above two estimations are in good agreement.

In the setup shown in Fig.~\ref{Fig1} at the location of the BEC we assume that the $y$ and $z$ components of the magnetic field of the CPW are negligible. This is a valid assumption for clouds that are much smaller than the dimensions of the cavity. Furthermore, we will approximate the amplitude of $B_{x}$ with its value at the center of the BEC, and neglect its spatial variation over the extension of the cloud. This assumption may be applied if the linewidth of the CPW resonator mode is much smaller than the bandwidth of the BEC and thus the resonance condition restricts the outcoupling to a thin disc around the center of the BEC.

The distance $d$ between the two systems has to be chosen in a way to maximize the coupling while avoiding the effect of van der Waals forces \cite{Schneeweiss2012,Jetter2013} between the trapped atoms and the CPW ($d_{0}\gtrapprox 1$ $\mu$m). At a distance of $d>(d_{0}+a)$ in the negative $y\rq$ direction, i.e., at the center of the BEC ($y=0$), according to Eqs.~(\ref{Bxyz}) the magnitude of the magnetic field is decreased by the factor $e^{-d\gamma_{1}}\approx e^{-\frac{\pi d}{b}}$, with the assumption that $\gamma_{1}\approx \pi/b$, which is valid for $b\ll \lambda$. Therefore, we estimate the amplitude of the magnetic field to be
\begin{equation}
B_{x}=e^{-\frac{\pi d}{b}}\left|B_{\mathrm{max}}\right|.
\end{equation}

In order to determine the density of outcoupled atoms at a spatial location $\rr$ due to the interaction with a single microwave photon  we use Eq.~(\ref{N_D})
\begin{equation}
N(\rr)= N(\Delta\approx 0,\rr)=\left|\dfrac{\hbar\eta}{Mgl_{0}}\right|^{2}D(0,\rr).
\end{equation}
In an experimental situation, the outcoupled atoms can be counted by ionizing a part of the beam of atoms with lasers. This defines a detection volume $V_{\mathrm{d}}$. To get the number of outcoupled atoms per unit time in this volume, we need to integrate $N(\rr)$ over $V_{\mathrm{d}}$
\begin{equation}
{\cal N}=\dfrac{N_{0}}{V_{\mathrm{BEC}}}\, \left(\frac{\sqrt{15}\mu_{B}}{8Mgl_{0}}\right)^{\!\!2}B_{x}^{2} \int_{V_{\mathrm{d}}}\overline{D(0,\rr)} d^{3}\rr,
\label{N_outcoup}
\end{equation}
where  $\overline{D(0,\rr)}=D(0,\rr)/(\mu/Ng_{s})$ is the dimensionless spectral resolution function (\ref{spec_res_func}) and we used the fact that the chemical potential of a condensate can be written as $\mu=5N_{0}g_{\mathrm{s}}/(2V_{\mathrm{BEC}})$ in the Thomas-Fermi approximation. In the following we will assume that the ionization detection has unit efficiency, i.e., the number of atoms in the above formula can be considered the detection signal in a unit of time. 

\subsection{Numerical estimation}
\label{subsec:num_at}

In order to get a numerical estimation of the number of outcoupled atoms by a single microwave photon let us consider a spherical condensate of radius $a=5$ $\mu$m with $N_{0}=2\times 10^{4}$ $^{87}$Rb atoms, which corresponds to a trapping frequency $\omega_{y}=2\pi\times 84$ Hz and a chemical potential $\mu/\hbar=2\pi \times 0.75$ kHz. The hyperfine splitting of the atoms is $\Omega\approx 2\pi \times 6.8347$ GHz and in an offset magnetic field $B_{0}=0.1$ mT in the $z$ direction the Zeeman splitting of the magnetic sublevels is $\omega_{0}=2\pi \times 0.7$ MHz, and therefore the frequency of the cavity needs to be $\omega_{\mathrm{CPW}}=2\pi \times 6.8354$ GHz. 

We assume that the CPW is constituted by Al-film conductors on a Sapphire substrate ($\varepsilon=11.5$) with dimensions $S=15$ $\mu$m, $W=10$ $\mu$m, $w=S/2$, and $h=500$ $\mu$m. Such CPW's have been measured to have a quality factor of $Q=1.72\times 10^{6}$ for low power, i.e., $\left<n_{\mathrm{photon}}\right>\approx 1$ \cite{Megrant2012}. The effective permittivity is determined to be $\varepsilon_{\mathrm{eff}}=6.25$, and the length of the cavity is $L=8.778$ mm.  Then, with a distance of $d=5$ $\mu$m between the center of the BEC and the CPW, the amplitude of the outcoupling magnetic field can be approximated to be $B_{x}\approx2.56$ nT. (Note that for the same parameters, the estimation based on the circuit QED result gives $\left|B^{\mathrm{CQED}}_{\mathrm{max}}\right|\approx 2.25$ nT). Based on our numerical results we find that by choosing the center of the detection volume at $\sim 65$ $\mu$m below the center of the condensate with a height of $y_{g}=60$ $\mu$m, the number of atoms per unit time in this volume is ${\cal N}\approx3$. Assuming ionization detection scheme with efficiency close to 1  \cite{Stibor2010}, all these atoms are detected, and ${\cal N}\approx 3$ calibrates the signal corresponding to the detection of a single microwave photon.

The value of ${\cal N}$ in our setup is mainly affected by the magnitude of the magnetic field, which can be increased by reducing the width $W$ between the center conductor and the ground planes of the CPW, but it would require to move the BEC closer to the cavity, which is also limited by the radius of the cloud. 
Increasing the number $N_{0}$ of trapped atoms with increasing cloud radius does not significantly increase $\cal N$, as can be seen from Eq.~(\ref{N_outcoup}). However increasing $N_{0}$ while keeping the cloud dimensions the same or enlarging the detection volume can obviously increase the number of outcoupled atoms.

\section{Conclusion}
\label{sec:concl}
We have evaluated the capabilities of a magnetically trapped Bose-Einstein condensate of Rubidium atoms to detect the magnetic field of a superconducting coplanar waveguide resonator by means of the atomlaser scheme. We have shown that by the counting of single atoms outcoupled by the measured field and falling out of the trap due to gravity, weak microwave fields at the single-photon level can be sensed and translated to a detectable signal of a few atoms with an integration time on the order of a second.

\begin{acknowledgments}
This work was supported by the National Research, Development and Innovation Office (Project Nos. K115624, K124351, PD120975, 2017-1.2.1-NKP-2017-00001). O. K. acknowledges support from the J\' anos Bolyai Research Scholarship of the Hungarian Academy of Sciences. 

\end{acknowledgments}

\appendix
\section{Effective dielectric constant and the capacitance of a CPW}
\label{AppA}

For a conventional CPW on a substrate of relative permittivity $\varepsilon_{r}$, the effective dielectric constant $\varepsilon_{\text{eff}}$ and the capacitance per unit length $c_{\mathrm{CPW}}$ can be analitically approximated using conformal mapping techniques as \cite{Simons_CPWbook}
\begin{align}
\varepsilon_{\mathrm{eff}}&=1+\frac{\varepsilon_{r}-1}{2}\frac{\kappa_{1}}{\kappa_{0}}, \label{eps_eff} \\
c_{\mathrm{CPW}}&=4\,\varepsilon_{0}\,\varepsilon_{\mathrm{eff}}\,\kappa_{0}, \label{c_CPW}
\end{align}   
where $\kappa_{i}=K\left(k_{i}\right)/K\left(k\rq_{i}\right)$, $K$ being the complete elliptic integral, $k_{0}=S/(S+2W)$, $k_{1}=\sinh\left(\frac{\pi S}{4 h}\right)/\sinh\left(\frac{\pi \left(S+2W\right)}{4 h}\right)$, $k\rq_{i}=\sqrt{1-k_{i}^{2}}$ for ($i=0,1$).

\section{Analytical formulas for the magnetic field of a CPW}
\label{AppB}

If the transverse size of the conductors and their distance is small relative to the mode wavelength of the cavity, then the quasi-static approximation may be applied \cite{Pozar_book}. Let us introduce the coordinates $x\rq=x$, $y\rq=y-d$, $z\rq=z+L/2$ (see Fig.~\ref{Fig1}) to describe the magnetic field components of the CPW, where $d$ is the distance between the center of the BEC and that of the CPW. For reasons of  symmetry, it suffices to restrict to one half of the structure (i.e., $0\leq x\rq\leq b$, $0\leq z\rq\leq L$), and consider the CPW as coupled slots \cite{Knorr1975}. We assume magnetic walls at $x\rq=0$ and $x\rq=b=S/2+W+w$ and open-circuit boundary conditions at $z\rq=0$ and $z\rq=L=\lambda_{g}/2$, with $\lambda_{g}$ being the mode wavelength \cite{Knorr1975,Simons1981}. As a result of the longitudinal boundary condition, the maxima of the electric and magnetic fields are shifted in space. The magnetic field is maximal at $z\rq=L/2$, where the electric field is zero, and zero at the two ends, where the electric field is maximal. Here we only present the components of the magnetic field for the case of odd (transverse) excitations, according to Ref. \onlinecite{Simons_CPWbook}. On the air side of the structure ($y\rq\leq0$) they can be approximated as
\begin{align}
B_{x}(\rr\rq)&\!=\! p \left[\sum\limits_{n>0}^{\infty}\frac{s_{n}}{F_{n}}\cos\!\left(\frac{n\pi x\rq}{b}\right)\!e^{-\gamma_{n}\left|y\rq\right|}\right]\!\! \sin\!\left(\frac{\pi z\rq}{L}\right)\!\!,  \notag \\
B_{y}(\rr\rq)&\!=\! p \left[\sum\limits_{n>0}^{\infty}s_{n}\sin\!\left(\frac{n\pi x\rq}{b}\right)\!e^{-\gamma_{n}\left|y\rq\right|}\right]\!\! \sin\!\left(\frac{\pi z\rq}{L}\right)\!\!, \\
B_{z}(\rr\rq)&\!=\! p \frac{2b}{\lambda_{g}}\left[\sum\limits_{n>0}^{\infty}q\frac{s_{n}}{n F_{n}}\sin\!\left(\frac{n\pi x\rq}{b}\right)\!e^{-\gamma_{n}\left|y\rq\right|}\right]\!\! \cos\!\left(\frac{\pi z\rq}{L}\right)\!\!, \notag
\label{Bxyz}
\end{align}
where 
\begin{align}
p=& -i \mu_{0}\mu_{r}\frac{4 V_{0}}{\eta b}\frac{\lambda}{\lambda_{g}},  \notag \\
s_{n}=&\dfrac{\sin\left(\frac{n\pi \delta}{2}\right)}{\frac{n\pi\delta}{2}}\sin\left(\frac{n\pi\bar{\delta}}{2}\right), \notag \\
q=&1-\left(\frac{\lambda}{\lambda_{g}}\right)^{2}, \\
F_{n}=&\frac{b\gamma_{n}}{n\pi}=\sqrt{1+\left(\frac{2bv}{n\lambda}\right)^{2}}, \notag 
\end{align}
with $\delta=W/b$, $\bar{\delta}=(S+W)/b$, $v=\sqrt{(\lambda/\lambda_{g})^{2}-1}$, $\mu_{0}$ is the permeability of free space, $\mu_{r}\approx1$ is the relative permittivity of the conductor, $\eta=\sqrt{\mu_{0}/\varepsilon_{0}}$ is the impedance of free space, and $V_{0}$ is the voltage directly across the slot between the central and ground electrodes. Let us note that the complex factor $i$ in the expression of $p$ is due to the fact that there is a $\pi/2$ phase difference in the time dependence of the magnetic field and the electric field, i.e., when the magnetic field is maximal in time, then the electric field is zero and vice versa. 

For the case when $b\ll \lambda, h$, where $h$ is the thickness of the substrate, it can be shown that the magnetic field on the substrate side of the structure is approximately given by
\begin{align}
B^{0\leq y\rq\leq h}_{x}(x\rq,y\rq,z\rq)\approx & -B_{x}(x\rq,-y\rq,z\rq), \notag \\
B^{0\leq y\rq\leq h}_{y}(x\rq,y\rq,z\rq)\approx & \,\, B_{y}(x\rq,-y\rq,z\rq), 
\end{align} 
while the $z$ components of the magnetic field can be neglected on both sides of the CPW, i.e., $\left|B_{z}(\rr\rq)\right|, |B^{0\leq y\rq\leq h}_{z}(\rr\rq)| \ll \left|B_{x}(\rr\rq)\right|, \left|B_{y}(\rr\rq)\right|$.

\section{Spectral resolution function of the BEC}
\label{AppC}

It was shown in Ref.~\onlinecite{Kalman2016}, that when the BEC is employed as a measuring device, its spectral resolution function can be written as
\begin{equation}
D(\omega-\Delta,\rr)=(Mgl_{0})^{2}\left|F(\omega-\Delta,\rr)\right|^{2}, \label{spec_res_func}
\end{equation}
where in the case of a spherical BEC with radus $a$, in the Tomas-Fermi approximation
\begin{widetext}
\begin{multline}
\!\!\!F(\omega-\Delta,\rr)=-\frac{\pi}{Mgl_{0}} \sqrt{\!\frac{\mu}{N\!g_{s}}}
\int\limits_{0}^{\infty}\!\!d\bar{k}_{\perp}\bar{k}_{\perp} 
 J_{0}(\bar{k}_{\perp}\bar{r}_{\perp}) \, \mathrm{Ci}\!\left(\bar{y}-\frac{E_{y}(\bar{k}_{\perp})}{Mgl_{0}}\right)
\int\limits_{0}^{1}\!\!d\bar{r}\rq_{\perp}\bar{r}\rq_{\perp} 
J_{0}(\bar{k}_{\perp}\bar{r}\rq_{\perp}) \\
\times
\int\limits_{-\bar{a}\sqrt{1-\bar{r}\rq^{2}_{\perp}}}^{\bar{a}\sqrt{1-\bar{r}\rq^{2}_{\perp}}}
d\bar{y}\rq 
\sqrt{1-\bar{r}\rq^{2}_{\perp}-\frac{\bar{y}\rq^{2}}{\bar{a}^{2}}}
\mathrm{Ai}\!\left(\bar{y}\rq-\frac{E_{y}(\bar{k}_{\perp})}{Mgl_{0}}\right).
\end{multline}
\end{widetext}
Here $\mathrm{Ci}$ is the complex Airy function $\mathrm{Ci}(x)=\mathrm{Bi}(x)+\mathrm{Ai}(x)$, $\bar{k}_{\perp}$ is the length of the dimensionless wave vector perpendicular to the direction of gravity, with $\bar{k}_{\perp}^{2}=a^{2}(k_{x}^{2}+k_{z}^{2})=2Ma^{2}(E_{x}+E_{z})/\hbar^{2}$, $E_{y}(\bar{k}_{\perp})=\hbar(\omega-\Delta)-\hbar^{2}\bar{k}_{\perp}^{2}/2Ma^{2}$, $\bar{r}\rq_{\perp}=\sqrt{x\rq^{2}+z\rq^{2}}/a$ is the length of the dimensionless position vector perpendicular to gravity, while $\bar{y}\rq=y/a$ is the dimensionless coordinate in the direction of gravity inside the  BEC. $\bar{r}_{\perp}$ refers to the dimensionless length of the position vector $\rr$ perpendicular to gravity.

\end{document}